\documentclass[12pt,reqno,titlepage]{article}
\usepackage{amsfonts,verbatim,graphicx}
\usepackage{amsmath}
\usepackage{ae,lscape}
\usepackage[
backend=biber,
style=apa,
sorting=nyt
]{biblatex}
\DeclareSourcemap{
  \maps[datatype=bibtex]{
    \map{
      \step[fieldset=issn, null]
      \step[fieldset=doi, null]
      \step[fieldset=url, null]
      \step[fieldset=urldate, null]
    }
  }
}

\usepackage{hyperref}
\hypersetup{
    colorlinks=true,
    urlcolor=blue,
    linkcolor=blue,
    citecolor=blue
}

 \providecommand{\possessivecite}[1]{\citeauthor{#1}'s\nolinebreak[2]
   \citeyearpar{#1}}

\newcommand{\abs}[1]{ \left | #1 \right | }

\renewcommand{\Re}{  \mathbf{R}  }

\newtheorem{thm}{Theorem}
\newtheorem{obs}[thm]{Observation}

\newtheorem{prop}[thm]{Proposition}
\newtheorem{ex}[thm]{Example}
\newtheorem{pb}[thm]{Problem}
\newtheorem{defn}[thm]{Definition}

\newenvironment{prf}[1][Proof]{\noindent\textbf{#1.} }{\ \rule{0.5em}{0.5em}}
\newcommand{\comments}[1]{}
\numberwithin{equation}{section}
\newenvironment{proof}[1][Proof]{\noindent\textbf{#1.} }{\ \rule{0.5em}{0.5em}}

\def\pw{\mathbin{>_j}}
\def\pm{\mathbin{>_i}}
\renewcommand{\P}[1]{\mathbin{>_{#1}}}

\input{tcilatex}
\begin{document}

\title{\textbf{Ordinal and cardinal solution concepts for two-sided matching}%
\thanks{%
The authors wish to thank Juan Pereyra Barreiro and Aditya Kuvalekar for
many useful suggestions, and for pointing out a mistake in one of the
examples in a previous version of the paper.} }
\author{\textbf{Federico Echenique}\thanks{%
Address: Division of the Humanities and Social Sciences, Mail Code 228-77,
Caltech, Pasadena, CA 91125, USA. E-mail: fede@caltech.edu.} \\
California Institute of Technology \and \textbf{Alfred Galichon}\thanks{%
Address: Department of Economics, Sciences Po, 75007 Paris, France. E-mail:
alfred.galichon@sciences-po.fr. This research has received funding from the
European Research Council under the European Union's Seventh Framework
Programme (FP7/2007-2013) / ERC grant agreement n%
${{}^\circ}$%
313699.} \\
Sciences Po, Paris}
\date{April 23, 2014}
\maketitle

\begin{abstract}
We characterize solutions for two-sided matching, both in the transferable and in the nontransferable-utility frameworks, using a cardinal formulation. Our approach makes the comparison of the matching models with and without transfers particularly transparent. We introduce the concept of a no-trade matching to study the role of transfers in matching. A no-trade matching is one in which the availability of transfers do not affect the outcome.\\
\bigskip

{\small

}
\end{abstract}

\newpage

\section{Introduction}
\setcounter{page}{1}
\setcounter{equation}{0}
We explore the role of transfers and cardinal utility in matching markets.
Economists regularly use one- and two-sided models, with and without
transfers. For example auctions allow for monetary transfers among the
agents, while models of marriage, organ donation and \textquotedblleft
housing\textquotedblright\ exchanges do not. There are two-sided matching
models of the labor market without transfers, such as the market for medical
interns in the US; and traditional models of the labor market where
salaries, and therefore transfers, are allowed. We seek to understand how
and why transfers matter in markets for discrete goods.

The question is interesting to us as theorists, but it also matters greatly
for one of the most important applications of matching markets, namely the
medical interns market. In the market for medical interns in the US (see 
\cite{roth84b}, \cite{roth90}, and \cite{roth2002}), hospitals match with
applicants using a centralized clearinghouse that implements a stable
matching. We always think of this market as one \textit{without} transfers,
because salaries are fixed first, before the matching is established. So at
the stage in which the parties ``bargain'' over who is to be matched to
whom, salaries are already fixed, and thus there are no transfers.

There is a priori no reason for things to be this way. Hospitals and interns
could instead bargain over salaries and employment at the same time. This is
arguably the normal state of affairs in most other labor markets; and it has
been specifically advocated for the medical interns market in the US (see %
\cite{crawford2008flexible}). It is therefore important to understand the
impact of disallowing transfers in a matching market. Our paper is a first
step towards understanding this problem.

In a two-sided matching market---for our purposes, in the Gale-Shapley
marriage market---this impact is important. We consider two canonical
models: the marriage market without transfers (the NTU model) and the
marriage market with transfers (the TU model, also called the assignment
game).

There are Pareto efficient, and even stable, matchings in the NTU model that
a utilitarian social planner would never choose, regardless of how she
weights agents' utilities. A utilitarian social planner has implicitly
access to transfers. Our results motivate an investigation into the distance
between the utilitarian welfare in the presence of transfers, and the
utilitarian welfare in the absence of transfers. We show that this gap can
be arbitrarily large. In fact, it can grow exponentially with the size of
the market. Of course this is established in an environment in which
utilities are bounded, and the bound is constant while the market grows
(otherwise the exercise would be meaningless). See our Proposition~\ref%
{prop:POA}\footnote{%
From the viewpoint of the recent literature in computer science on the
\textquotedblleft price of anarchy\textquotedblright\ (see e.g.\ %
\cite{roughgarden2005selfish}), Proposition~\ref{prop:POA} says that the
\textquotedblleft price of no transfers\textquotedblright\ can be
arbitrarily bad, and grow super-exponentially with the size of the market.}.

We present results characterizing Pareto efficiency and the role of
transfers in marriage models. Ex-ante Pareto optimality in the model with
transfers is characterized by the maximization of the weighted utilitarian
sum of utilities, while Pareto optimality when there are no transfers is
equivalent to a different maximization problem, one where the weighted sum
of ``adjusted'' utilities are employed. Each of these problems, in turn,
have a formulation as a system of linear inequalities. The results follow
(perhaps unexpectedly) from Afriat's theorem in the theory of revealed
preference.

In order to explore the role of transfers, we study a special kind of stable
matching: A \emph{no-trade stable matching} in a marriage market is a
matching that is not affected by the presence of transfers. This is the
central notion in our paper. Agents are happy remaining matched as specified
by the matching, even if transfers are available, and \emph{even though they
do not make use of transfers.} Transfers are available, but they are not
needed to support the stable matching. There is thus a clear sense in which
transfers play no role in a no-trade stable matching.

The notion of no-trade stable matching is useful for the following reason.
We can think of transfers as making some agents better off at the expense of
others. It is then possible to modify a market by choosing a cardinal
utility representation of agents preferences with the property that the
matching remains stable with and without transfers (Theorem~\ref%
{thm:onematching}). Under certain circumstances, namely when the stable
matchings are \textquotedblleft isolated,\textquotedblright\ we can choose a
cardinal representation that will work in this way for every stable
matching. So under such a cardinal representation of preferences, any stable
matching remains stable regardless of the presence of transfers. Finally
(Example~\ref{ex:negative}), we cannot replicate the role of transfers by
re-weighting agents' utilities. In general, to instate a no-trade stable
matching, we need the full freedom of choosing alternative cardinal
representations.

It is easy to generate examples of stable matchings that cannot be sustained
when transfers are allowed, and of stable matchings that can be sustained
with transfers (in the sense of being utilitarian-efficient), but where
transfers are actually used to sustain stability. We present conditions
under which a market has a cardinal utility representation for which stable
matchings are no trade matchings.

In sum, the notion of a no-trade stable matching captures both TU and NTU
stability: a no-trade stable matching is also a TU and NTU stable matching.
TU stability is, on the other hand, strictly stronger than ex-ante Pareto
efficiency, which is strictly stronger than ex-post Pareto efficiency. NTU
stability is strictly stronger than ex-post Pareto efficiency.\footnote{%
TU and NTU stability are not comparable in this sense. Empirically, though,
they are comparable, with TU stability having strictly more testable
implications than NTU stability \cite{echleeshumyen11}.}

The model without transfers was introduced by \cite{gale62}. The model with
transfers is due to \cite{shapl71}. \cite{kelso82} extended the models
further, and in some sense Kelso and Crawford's is the first paper to
investigate the effects of adding transfers to the Gale-Shapley marriage
model. \cite{roth84} and \cite{hatfi04} extended the model to allow for
more complicated contracts, not only transfers (see \cite{hatfi08} and %
\cite{echen12} for a discussion of the added generality of contracts). We
are apparently the first to consider the effect of transfers on a given
market, with specified cardinal utilities, and the first to study the notion
of a no-trade stable matching.

\section{The Marriage Problem}

\subsection{The model}

Let $M$ and $W$ be finite and disjoint sets of, respectively, men and women,
which are assumed to be in equal number; $M\cup W$ comprise the \textit{%
agents} in our model. We can formalize the marriage \textquotedblleft
market\textquotedblright\ of $M$ and $W$ in two ways, depending on whether
we assume that agents preferences have cardinal content, or that they are
purely ordinal. For our results, it will be crucial to keep in mind the
difference between the two frameworks.

An \emph{ordinal marriage market} is a tuple $\left( M,W,P\right) $, where $%
P $ is a \emph{preference profile}: a list of preferences $>_{i}$ for every
man $i$ and $>_{j}$ for every woman $j$. Each $>_{i}$ is a linear order over
$W$, and each $>_{j}$ is a linear order over $M$. Here, agents always prefer
being matched with anyone rather than being unmatched. The weak order
associated with $>_{s}$ is denoted by $\geq _{s}$ for any $s\in M\cup W$.%
\footnote{%
A linear order is a binary relation that is complete, transitive and
antisymmetric. The weak order $\geq _{s}$ is defined as $a\geq _{s}b$ if $%
a=b $ or if $a>_{s}b$.}

We often specify a preference profile by describing instead utility
functions for all the agents. A \emph{cardinal marriage market} is a tuple $%
\left( M,W,U,V\right) $, where $U$ and $V$ define the agents' utility
functions: $U\left( i,j\right) $ (resp.\ $V\left( i,j\right) $) is the
amount utility derived by man $i$ (resp.\ woman $j$) out of his match with
woman $j$ (resp.\ man $i$). The utility functions $U$ and $V$ \emph{represent%
} $P$ if, for any $i$ and $i^{\prime }$ in $M$, and $j$ and $j^{\prime }$ in
$W$,
\begin{eqnarray*}
U(i,j) &>&U(i,j^{\prime })\iff j>_{i}j^{\prime }\text{, and} \\
V(i,j) &>&V(i^{\prime },j)\iff i>_{j}i^{\prime }.
\end{eqnarray*}

We say that $U$ and $V$ are a \emph{cardinal representation of $P$}.
Clearly, for any cardinal marriage market $\left( M,W,U,V\right) $ there is
a corresponding ordinal market.

A one-to-one function $\sigma :M\rightarrow W$ is called a \textit{matching}%
. When $w=\sigma (m)$ we say that $m$ and $w$ are matched, or married, under
$\sigma $. In our setting, under a given matching, each man or woman is
married to one and only one partner of the opposite sex. We shall denote by $%
\mathcal{A}$ the set of matchings. We shall assume that $M$ and $W$ have the
same number of elements, so that $\mathcal{A}$ is non-empty.

In our definition of matching is that agents are always married: we do not
allow for the possibility of singles. 

Under our assumptions, we can write $M=\{m_1,\ldots, m_n\}$ and $%
W=\{w_1,\ldots, w_n\}$. For notational convenience, we often identify $m_i$
and $w_j$ with the numbers $i$ and $j$, respectively. So when we write $%
j=\sigma(i)$ we mean that woman $w_j$ and man $m_i$ are matched under $%
\sigma $. This usage is a bit different than what it standard notation in
matching theory, but it makes the exposition of our results a lot simpler.

We shall often fix an arbitrary matching, and without loss of generality let
this matching be the \textit{identity matching}, denoted by $\sigma_0$. That
is,
\begin{equation*}
\sigma _{0}\left( i\right) =i.
\end{equation*}

For a matching $\sigma $, let $u_{\sigma }(i)=U(i,\sigma (i))$ and $%
v_{\sigma }(j)=V(\sigma ^{-1}(j),j)$. When $\sigma =\sigma _{0}$, we shall
often omit it as a subscript and just use the notation $u$ and $v$.

One final concept relates to random matchings. We consider the possibility
that matching is chosen according to a lottery: a \emph{fractional matching}
is a matrix $\pi =(\pi _{i,j})$ such that $\pi _{ij}\geq 0$ and letting $\pi
_{ij}$ the probability that individuals $i$ and $j$ get matched, the
constraints on $\pi $ are
\begin{equation*}
1=\sum_{i^{\prime }=1}^{n}\pi _{i^{\prime }j}=\sum_{j^{\prime }=1}^{n}\pi
_{ij^{\prime }}\text{ }\forall i,j\in \left\{ 1,...,n\right\}
\end{equation*}%
(i.e.\ $\pi $ is a \emph{bistochastic matrix}). It is a celebrated result
(the Birkhoff von-Neumann Theorem) that such matrices result from a lottery
over matchings. Let $\mathcal{B}$ denote the set of all fractional matchings.

\subsection{Solution concepts}

\label{sec:solutions}

We describe here some commonly used solution concepts. The first solutions
capture the notion of Pareto efficiency. In second place, we turn to notion
of core stability for matching markets. For simplicity of exposition, we
write these definitions for the specific matching $\sigma _{0}$. Of course
by relabeling we can express the same definitions for an arbitrary matching.

A solution concept singles out certain matchings as immune to certain
alternative outcomes that could be better for the agents. If we view such
alternatives as arising ex-post, after any uncertainty over which matching
arises has been resolved, then we obtain a different solution concept than
if we view the alternatives in an ex-ante sense.

\subsubsection{NTU Pareto efficiency}

Matching $\sigma _{0}\left( i\right) =i$ is \emph{ex-post NTU Pareto
efficient}, or simply \emph{ex-post Pareto efficient} if there is no
matching $\sigma $ that is at least as good as $\sigma _{0}$ for all agents,
and strictly better for some agents. That is, such that the inequalities $%
U\left( i,\sigma \left( i\right) \right) \geq U\left( i,i\right) $ and $%
V\left( \sigma ^{-1}\left( j\right) ,j\right) \geq V\left( j,j\right) $
cannot simultaneously hold with at least one strict inequality.

In considering alternative matchings, it is easy to see that one can
restrict oneself to cycles. The resulting formulation of efficiency is very
useful, as it allows us to relate efficiency with standard notions in the
literature on revealed preference.

Hence matching $\sigma _{0}\left( i\right) =i$ is ex-post Pareto efficient
if and only if for every cycle\footnote{%
Recall that a cycle $i_{1},...,i_{p},i_{p+1}=i_{1}$ is a permutation $\sigma
$ such that $\sigma \left( i_{1}\right) =i_{2}$, $\sigma \left( i_{2}\right)
=i_{3}$,...,$\sigma \left( i_{p-1}\right) =i_{p}$, and $\sigma \left(
i_{p}\right) =i_{1}$.}\label{footnotecyclepage} $i_{1},...,i_{p+1}=i_{1}$,
inequalities $U\left( i_{k},i_{k+1}\right) \geq U\left( i_{k},i_{k}\right) $
and $V\left( i_{k},i_{k+1}\right) \geq V\left( i_{k},i_{k}\right) $ cannot
hold simultaneously unless they all are equalities. In other words:

\begin{obs}
Matching $\sigma _{0}\left( i\right) =i$ is ex-post Pareto efficient if for
every cycle $i_{1},...,i_{p+1}=i_{1}$, and for all $k$, inequalities%
\begin{equation*}
U\left( i_{k},i_{k+1}\right) \geq U\left( i_{k},i_{k}\right) \text{ and }%
V\left( i_{k},i_{k+1}\right) \geq V\left( i_{k},i_{k}\right) \text{,}
\end{equation*}%
cannot hold simultaneously unless they are all equalities.
\end{obs}

In an ex-ante setting, we can think of probabilistic alternatives to $\sigma
_{0}$. As a result, we obtain the notion of ex-ante Pareto efficiency. To
define this notion, we require not only that there is no other matching
which is preferred by every individual, but also that there is no lottery
over matchings that would be preferred.

Formally: Matching $\sigma _{0}\left( i\right) =i$ is \emph{ex-ante NTU
Pareto efficient} or simply \emph{ex-ante Pareto efficient}, if for any $\pi
\in \mathcal{B}$, and for all $i$ and $j$, inequalities%
\begin{equation*}
\sum_{j}\pi _{ij}U\left( i,j\right) \geq U\left( i,i\right) \text{ and }%
\sum_{i}\pi _{ij}V\left( i,j\right) \geq V\left( j,j\right)
\end{equation*}%
cannot hold simultaneously unless they are all equalities.

Note that the problem of ex-post efficiency is purely ordinal, as ex-post
efficiency of some outcome only depends on the rank order preferences, not
on the particular cardinal representation of it. In contrast, the problem of
ex-ante efficiency is cardinal, as we are adding and comparing utility
levels across states of the world. The concept of NTU\ Pareto efficiency is
interesting for example in the context of school choice problems, where the
assignment is often not deterministic, and transfers are not permitted. In
this context, this concept provides an adequate assessment of welfare.

\subsubsection{TU\ Pareto efficiency}

We now assume that utility is transferable across individuals. In this case,
a matching is Pareto efficient if no other matching produces a higher
welfare, accounted for as the sum of individual cardinal utilities. It is a
direct consequence of the Birkhoff-von Neumann theorem that if a fractional
matching produces a higher welfare, then some deterministic matching also
produces a higher welfare. As a result, the notions of ex-ante and ex-post
TU Pareto efficiency coincide, and we do not need to distinguish between
them.

Matching $\sigma _{0}\left( i\right) =i$ is \emph{TU Pareto efficient} if
there is no matching $\sigma $ for which
\begin{equation*}
\sum_{i=1}^{n}U(i,\sigma (i))+V(i,\sigma (i))>\sum_{i=1}^{n}\left(
U(i,i)+V(i,i)\right) .
\end{equation*}

\begin{obs}
Matching $\sigma _{0}\left( i\right) =i$ is \emph{TU Pareto efficient} if
for every cycle $i_{1},...,i_{p+1}=i_{1}$, and for all $k$, inequalities%
\begin{equation*}
\sum_{k=1}^{p}U\left( i_{k},i_{k+1}\right) +V\left( i_{k},i_{k+1}\right)
\geq \sum_{k=1}^{p}U\left( i_{k},i_{k}\right) +V\left( i_{k+1},i_{k+1}\right)
\end{equation*}%
cannot hold simultaneously unless they are all equalities.
\end{obs}

In the previous definitions, transfers are allowed across any individuals.
One may have considered the possibility of transfers only between matched
individuals. It is however well known since \cite{shapl71} that this
apparently more restrictive setting leads in fact to the same notion of
efficiency. As we recall below, TU\ Pareto efficiency is equivalent to TU
stability, so to avoid confusions, we shall systematically refer to
\textquotedblleft TU\ stability\textquotedblright\ instead of
\textquotedblleft TU Pareto efficiency,\textquotedblright\ and, in the
sequel, reserve the notion of efficiency for NTU\ efficiency.

\subsubsection{NTU\ Stability}

We now review notions of stability. Instead of focusing on the existence of
a matching which would be an improvement for everyone (as in Pareto
efficiency), we focus on matchings which would be an improvement for a newly
matched pair of man and woman. Thus we obtain two solution concepts,
depending on whether we allow for transferable utility.

Our definitions are classical and trace back to \cite{gale62} and %
\cite{shapl71}. See \cite{roth90} for an exposition of the relevant theory.

Matching $\sigma _{0}\left( i\right) =i$ is \emph{\ stable in the
nontransferable utility matching market}, or \emph{NTU\ stable} if there is
no \textquotedblleft blocking pair\textquotedblright\ $\left( i,j\right) $,
i.e.\ a pair $\left( i,j\right) $ such that $U\left( i,j\right) >U\left(
i,i\right) $ and $V\left( i,j\right) >V\left( j,j\right) $ simultaneously
hold.

Hence, using our assumptions on utility, we obtain the following:

\begin{defn}
Matching $\sigma _{0}\left( i\right) =i$ is NTU stable if%
\begin{equation*}
\forall i,j:~\min \left( U\left( i,j\right) -U\left( i,i\right) ,V\left(
i,j\right) -V\left( j,j\right) \right) \leq 0.
\end{equation*}
\end{defn}

Of course, this notion is an ordinal notion and should not depend on the
cardinal representation of men and women's preferences, only on the
underlying ordinal matching market.

\subsubsection{TU Stability}

Utility is transferable across pair $\left( i,j\right) $ if there is the
possibility of a utility transfer $t$ (of either sign) from $j$ to $i$ such
that the utility of $i$ becomes $U\left( i,j\right) +t$, and utility of $j$
becomes $V\left( i,j\right) -t$. When we assume that utility is
transferable, in contrast, we must allow blocking pairs to use transfers.
Then a couple $(i,j)$ can share, using transfers, the \textquotedblleft
surplus\textquotedblright\ $U(i,j)+V(i,j)$. Thus we obtain the definition:

\begin{defn}
\label{def:TUStable}Matching $\sigma _{0}\left( i\right) =i$ is \emph{TU\
stable}, if there are vectors $\tilde{u}\left( i\right) $ and $\tilde{v}%
\left( j\right) $ such that for each $i$ and $j$,%
\begin{equation*}
\tilde{u}\left( i\right) +\tilde{v}\left( j\right) \geq U\left( i,j\right)
+V\left( i,j\right)
\end{equation*}%
must hold with equality for $i=j$.
\end{defn}

By a celebrated result of \cite{shapl71}, this notion is equivalent to the
notion of TU Pareto efficiency. Note that there may be multiple vectors $%
\tilde u$ and $\tilde v$ for the given matching $\sigma_0$.

\subsection{No-Trade stability}

The notions of TU and NTU stability have been known and studied for a very
long time. Here, we seek to better understand the effect that the
possibility of transfers has on a matching market. We introduce a solution
concept that is meant to relate the two notions.

Note that if matching $\sigma _{0}\left( i\right) =i$ is TU stable, then
there are transfers between the matched partners, say from woman $i$ to man $%
i$, equal to
\begin{equation}
T_{i}= \tilde u\left( i\right) -U\left( i,i\right) =V\left( i,i\right)
-\tilde v\left( i\right)  \label{eq:transfersTU}
\end{equation}%
where the payoffs $\tilde u\left( i\right) $ and $\tilde v\left( j\right) $
are those of Definition~\ref{def:TUStable}. We want to understand the
situations when matching $\sigma _{0}\left( i\right) =i$ is TU stable but
when no actual transfers are made ``in equilibrium.'' As a result, the
matching $\sigma _{0}$ is NTU stable as well as it is TU stable.

\bigskip

We motivate the notion of a No-Trade stable matching with an example. We
present a matching market with a matching which is both the unique TU stable
matching and also the unique NTU stable matching. In order for agents to
\emph{accept it,} however, transfers are needed.

\begin{ex}
In this and other examples, we write the payoffs $U$ and $V$ in matrix form.
In the matrices, the payoff in row $i$ and column $j$ is the utility $U(i,j)$
for man $i$ in matrix $U$, and utility $V(i,j)$ for woman $j$ in matrix $V$.

\label{ex:notNTM} Consider the following utilities
\begin{equation*}
U=\left(
\begin{array}{ccc}
0 & 2 & 1 \\
1 & 2 & 0 \\
1 & 0 & 2%
\end{array}%
\right) ,~V=\left(
\begin{array}{ccc}
2 & 1 & 0 \\
1 & 2 & 0 \\
0 & 1 & 2%
\end{array}%
\right)
\end{equation*}

Note that the matching $\sigma _{0}(i)=i$ is the unique NTU\ stable
matching, and is also the unique TU stable matching. To sustain it in the TU
game, however, requires transfers. Indeed, $u(i_{1})=0$ and $v(j_{2})=2$
cannot be TU stable payoffs as
\begin{equation*}
2=u(i_{1})+v(j_{2})<U(i_{1},j_{2})+V(i_{1},j_{2})=3
\end{equation*}%
contradicts Definition~\ref{def:TUStable}. Intuitively, one needs to
compensate agent $i=1$ in order for him to remained matched with $j=1$.
Hence, even though $\sigma _{0}$ is NTU-stable and TU-stable, transfers
between the agents are required to \emph{sustain} it as TU-stable.
Anticipating the definition to follow, this means this matching is not a
No-trade stable matching.
\end{ex}

Matching $\sigma _{0}\left( i\right) =i$ is \emph{no-trade stable} when it
is TU stable and there are no actual transfers between partners at
equilibrium. In other words, Equation~\eqref{eq:transfersTU} should hold
with $T_{i}=0$. That is, $U\left( i,i\right) =u\left( i\right) $, $V\left(
j,j\right) =v\left( j\right) $, and so:

\begin{defn}[No-Trade Matching]
Matching $\sigma _{0}\left( i\right) =i$ is no-trade stable if and only if
for all $i$ and $j$,%
\begin{equation*}
U\left( i,j\right) +V\left( i,j\right) \leq U\left( i,i\right) +V\left(
j,j\right) .
\end{equation*}
\end{defn}

Therefore in a no-trade\ stable matching, two matched individuals would have
the opportunity to operate monetary transfers, but they choose not to do so.
To put this in different terms, in a no-trade\ stable matching, spouses are
\textquotedblleft uncorrupted\textquotedblright\ because no monetary
transfer actually takes place between them, but they are not
\textquotedblleft incorruptible\textquotedblright , because the rules of the
game would allow for it.

\section{Cardinal characterizations}

\label{sec:MainResults}

We now present simple characterizations of the solution concepts described
in Section~\ref{sec:solutions}.

Our characterizations involve cardinal notions, even for the solutions that
are purely ordinal in nature. The point is to characterize all solutions
using similar concepts, so it is easier to understand how the solutions
differ. It will also help us understand the role of transfers in matching
markets.

Define $\sigma _{0}$ as the matching such that $\sigma _{0}(i)=i$. We need
to introduce the following notation:
\begin{eqnarray*}
R_{ij} &=&U\left( i,i\right) -U\left( i,j\right) \\
S_{ij} &=&V\left( j,j\right) -V\left( i,j\right) ,
\end{eqnarray*}%
defined for each $i\in M$ and $j\in W$. Note that $R_{ij}$ measures how much
$i$ prefers his current partner to $j$, and $S_{ij}$ measures how much $j$
prefers her current partner to $i$. These two concepts are dependent on the
matching $\sigma _{0}$, which we take as fixed in the following result.

\begin{thm}
\label{prop:EfficiencyMarriage} Matching $\sigma _{0}\left( i\right) =i$ is:

(a) No-trade stable iff for all $i$ and $j$ in $\left\{ 1,...,n\right\} $%
\begin{equation}
0\leq R_{ij}+S_{ij}  \label{NTS}
\end{equation}

(b) NTU stable iff for all $i$ and $j$ in $\left\{ 1,...,n\right\} $
\begin{equation}
0\leq \max \left( R_{ij},S_{ij}\right)  \label{NTUStability}
\end{equation}

(c) TU stable iff there exists $T\in \mathbb{R}^{n}$ such that for all $i$
and $j$ in $\left\{ 1,...,n\right\} $
\begin{equation}
T_{j}-T_{i}\leq R_{ij}+S_{ij}  \label{TUStability}
\end{equation}

(d) Ex-ante Pareto efficient iff there exist $v_{i}$ and $\lambda _{i},\mu
_{j}>0$ such that for all $i$ and $j$ in $\left\{ 1,...,n\right\} $
\begin{equation}
v_{j}-v_{i}\leq \lambda _{i}R_{ij}+\mu _{j}S_{ij}  \label{EAP}
\end{equation}

(e) Ex-post Pareto iff there exist $v_{i}$ and $\lambda _{i}>0$ such that
for all $i$ and $j$ in $\left\{ 1,...,n\right\} $
\begin{equation}
v_{j}-v_{i}\leq \lambda _{i}\max \left( R_{ij},S_{ij}\right) .  \label{EPP}
\end{equation}
\end{thm}

Observe that ~\eqref{EAP} and~\eqref{EPP} are ``Afriat inequalities,'' using
the terminology in revealed preference theory.

As a consequence of the previous characterizations, it is straightforward to
list the chains of implications between the various solution concepts.

\begin{figure}[!ht]
\begin{center}
\includegraphics[width=10cm]{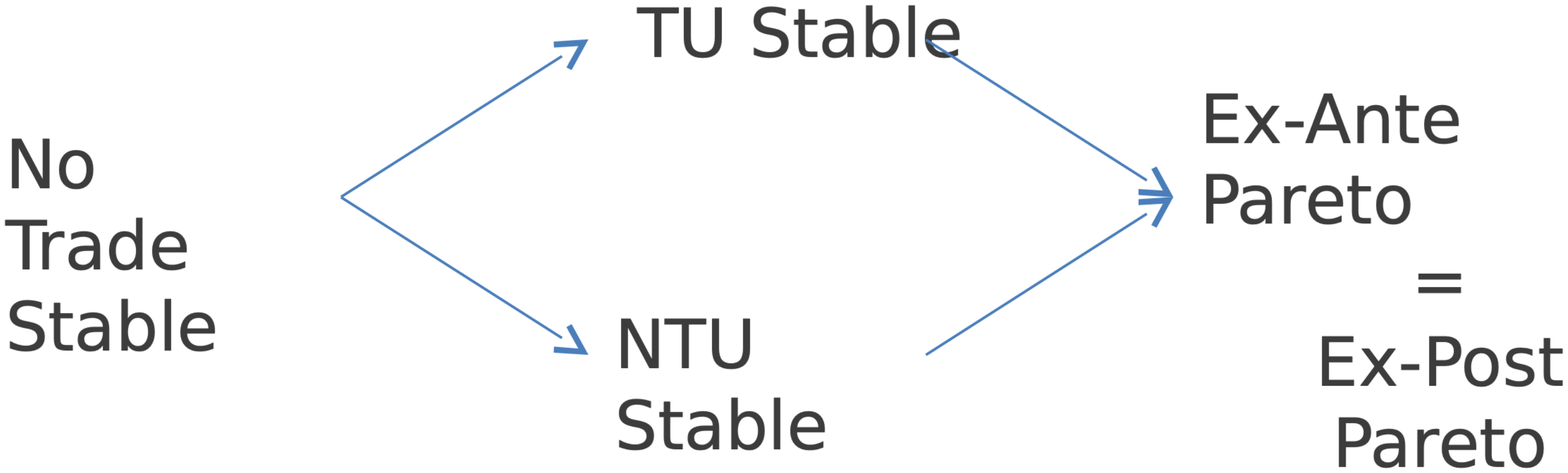}
\end{center}
\caption{Summary of the implications in Theorem \protect\ref%
{thm:Implications}, part (ii) when there are only two agents on each side of
the market.}
\label{fig2agents}
\end{figure}

\begin{figure}[!ht]
\begin{center}
\includegraphics[width=10cm]{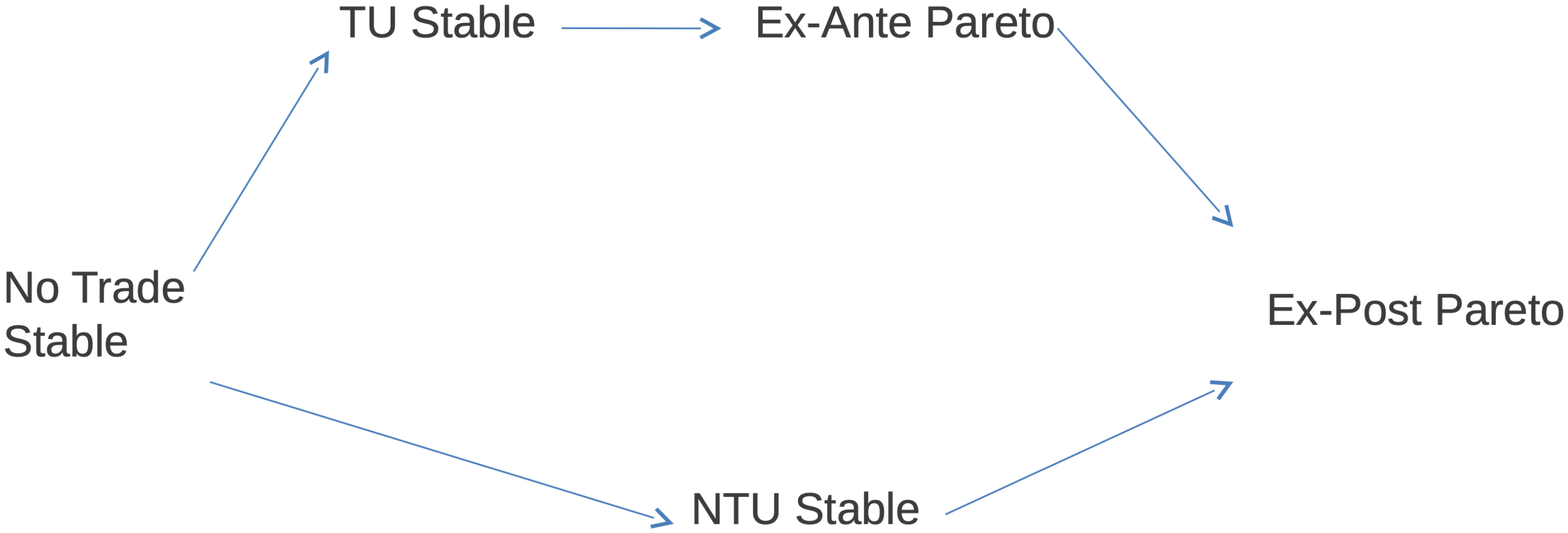}
\end{center}
\caption{Summary of the implication in Theorem \protect\ref{thm:Implications}%
, part (iii) when there are three agents or more on each side of the market.}
\label{fig:implications}
\end{figure}

\begin{thm}
\label{thm:Implications}(i) The two following chains of implications always
hold:

-- No-trade Stable implies TU Stable, TU Stable implies Ex-ante Pareto
Efficient, Ex-ante Pareto Efficient implies Ex-post Pareto Efficient, and

-- No-trade Stable implies NTU stable, NTU stable implies Ex-post Pareto
Efficient.

(ii) Assume that there are two agents on each side of the market. Then two
additional implications hold:

-- Ex-Post Pareto Efficient implies (and thus is equivalent to) Ex-Ante
Pareto Efficient, and

-- NTU\ stable implies Ex-Ante Pareto Efficient.

Any further implication which does not logically follow from those written
is false. See Figure (\ref{fig2agents}).

(iii) Assume that there are at least three agents on each side of the
market. Then any implication that does not logically follow from the ones
stated in part (i) of Theorem \ref{prop:EfficiencyMarriage} above are false.
See Figure (\ref{fig:implications}).
\end{thm}

The implications in Theorem~\ref{thm:Implications} are illustrated in
Figures (\ref{fig2agents}) and (\ref{fig:implications}).

The proof of Theorem~\ref{thm:Implications} is given in the Appendix. It
relies on the following counterexamples.

\begin{ex}
\label{ex1}Consider%
\begin{equation}
U=%
\begin{pmatrix}
0 & -2 \\
1 & 0%
\end{pmatrix}%
\text{ and }V=%
\begin{pmatrix}
0 & -2 \\
1 & 0%
\end{pmatrix}%
\end{equation}

Then $\sigma _{0}$ is TU\ stable (and thus Ex-Ante Pareto Efficient and
Ex-Post Pareto Efficient), but not NTU stable, and not No-trade stable.
\end{ex}

\bigskip

\begin{ex}
\label{ex2}Consider%
\begin{equation*}
U=\left(
\begin{array}{ccc}
0 & 2 & -1 \\
-1 & 0 & 2 \\
2 & -1 & 0%
\end{array}%
\right) ,~V=\left(
\begin{array}{ccc}
0 & -1 & 2 \\
2 & 0 & -1 \\
-1 & 2 & 0%
\end{array}%
\right)
\end{equation*}

Then $\sigma _{0}$ is NTU\ stable (hence Ex-Post Pareto efficient). But it
is not Ex-Ante Pareto efficient (hence neither No-trade stable nor TU
stable). Indeed consider the fair lottery over the 6 existing pure
assignments. Under this lottery, each agent achieves a payoff of 1/3, hence
this lottery is ex-ante preferred by each agent to $\sigma _{0}$.
\end{ex}

\bigskip

\begin{ex}
\label{ex3}Consider now
\begin{equation}
U=%
\begin{pmatrix}
0 & 1 \\
-2 & 0%
\end{pmatrix}%
\text{ and }V=%
\begin{pmatrix}
0 & 1 \\
3 & 0%
\end{pmatrix}
\label{UV2by2bis}
\end{equation}

Then $\sigma _{0}$ is ex-ante Pareto efficient, and it is ex-post Pareto
efficient, but it is not TU\ stable, and it is not NTU stable. Ex-ante
Pareto efficiency follows from $\lambda _{i}=5$, $\mu _{j}=1$ and $v_{2}=2$,
$v_{1}=1$. Ex-post Pareto efficiency is clear. It is easily seen that $%
\sigma _{0}$ is not TU stable. It is also clear that the matching is not NTU
stable, as $i=1$, $j=2$ form a blocking pair.
\end{ex}

\bigskip

\section{NTU Stability and No-trade stable matchings}

As we explained above, we use no-trade stability to shed light on the role
of transfers. Given a stable NTU\ matching, one may ask if is there a
cardinal representation of the agents' utility such that the stable matching
is no-trade stable.

As we shall see, the answer is yes if we are allowed to tailor the cardinal
representation to the given stable matching. If we instead want a
representation that works for all stable matchings in the market, we shall
resort to a regularity condition: that the matchings be isolated.

Finally, some statements in Theorem~\ref{prop:EfficiencyMarriage} involve
the rescaling of utilities: they show how optimality can be understood
through the existence of \emph{weights} on the agents that satisfy certain
properties. We can similarly imagine finding, not an arbitrary cardinal
representation of preferences, but a restricted rescaling of utilities that
captures the role of transfers. That is to say, a rescaling of utilities
that ensures that the matching is no-trade stable. We shall present an
example to the effect that such a rescaling is not possible.

\subsection{Linking models with and without transfers}

Our first question is whether for a given NTU stable matching, there is a
cardinal representation of preferences under which the same matching is
No-trade stable. The answer is yes.

\begin{thm}
\label{thm:onematching} Let $\left( M,W,P\right) $ be an ordinal matching
market. If $\sigma $ is a NTU stable matching, then there is a cardinal
representation of $P$ such that $\sigma $ is a no-trade stable matching in
the corresponding cardinal market.
\end{thm}

It is natural to try to strengthen this result in two directions. First, we
could expect to choose the cardinal representation of preferences as a
linear rescaling of a given cardinal representation of the preferences.
Given what we know about optimality being characterized by choosing
appropriate utility weights, it makes sense to ask whether any stable
matching can be obtained as a No-Trade Stable Matching if one only weights
agents in the right way. Namely:

\begin{pb}
Is it the case that matching $\sigma _{0}$ is NTU stable if and only if
there is $\lambda _{i},\mu _{j}>0$ such that for all $i$ and $j$,
\begin{equation*}
0\leq \lambda _{i}R_{ij}+\mu _{j}S_{ij}?
\end{equation*}
\end{pb}

After all, transfers favor some agents over others, and utility weights play
a similar role. Our next example shows that this is impossible. It exhibits
a stable matching that is not No-Trade for any choice of utility weights.

\begin{ex}
\label{ex:negative} Consider a marriage market defined as follows. The sets
of men and women are: $M=\left\{ m_{1},m_{2},m_{3},m_{4}\right\} $ and $%
W=\left\{ w_{1},w_{2},w_{3},w_{4}\right\} $. Agents' preferences are defined
through the following utility functions:
\begin{equation*}
U=\left(
\begin{array}{cccc}
1.01 & 0 & \underline{1}/2 & -1 \\
0 & 1 & -1 & \underline{1}/2 \\
\underline{1}/2 & 1/5 & 1/3 & 1/4 \\
1/5 & \underline{1}/2 & 1/3 & 1/4%
\end{array}%
\right) ,~V=\left(
\begin{array}{cccc}
0 & 1 & \underline{1}/2 & -1 \\
1 & 0 & -1 & \underline{1}/2 \\
\underline{1}/2 & 1/5 & 1/3 & 1/4 \\
1/5 & \underline{1}/2 & 1/3 & 1/4%
\end{array}%
\right)
\end{equation*}

The unique stable matching is underlined. Uniqueness is readily verified by
running the Gale-Shapley algorithm. So $u_{i}=v_{j}=1/2$ for all $i$ and $j$%
. Yet it is shown in the appendix that there are no $\lambda _{i},\mu _{j}>0$
such that for all $(i,j)$,
\begin{equation*}
\lambda _{i}(u_{i}-U_{ij})+\mu _{j}(v_{j}-V_{ij})\geq 0.
\end{equation*}

This example has indifference in payoffs. It is simple to perturb the
payoffs so that there are no more indifference, and the conclusion still
holds. This is explained in the proof in the appendix.
\end{ex}

Example~\ref{ex:negative} has the following implication (which also follows
from Example \ref{ex:notNTM}).

\begin{corollary}
There are cardinal matching markets that do not possess a no-trade matchings.
\end{corollary}

Given Example~\ref{ex:negative}, it is clear that No-Trade can only be
achieved by appropriate choice of agents' utility functions. Our next
question deals with the existence of cardinal utilities such that the set of
No-trade stable matchings and NTU stable matchings will coincide for \emph{%
all} stable matchings in a market. We show that if the stable matchings are
isolated then one can choose cardinal utilities such that all stable
matchings are No-Trade.

Let $S(P)$ denote the set of all stable matchings. A matching $\sigma\in
S(P) $ is \emph{isolated} if $\sigma^{\prime }(a)\neq \sigma (a)$ for all $%
a\in M\cup W$ and all $\sigma^{\prime }\in S(P)\setminus \{\sigma \}$.

\begin{thm}
\label{prop:isolated} There is a representation $(U,V)$ of $P$ such that for
all $\sigma \in S(P)$, if $\sigma $ is isolated then $\sigma$ is no trade
stable for $(U,V)$.
\end{thm}

The question whether the conclusion holds without the assumption that the
matching is isolated remains open to investigation.

\subsection{Price of no transfers}

The logic of the previous subsection can be pushed further, to obtain a
``Price of Anarchy,'' in the spirit of the recent literature in computer
science (\cite{roughgarden2005selfish}). We \emph{quantify} the cost in
social surplus (sum of agents' utilities) that results from NTU stability:
we can think of this cost as an efficiency gap inherent in the notion of NTU
stable matching. The result is that the gap can be arbitrarily large, and
that it grows ``super exponentially'' in the size of the market (i.e.\ it
grows at a faster rate than $n^g$, for any $g$, where $n$ is the size of the
market).

Let $\Delta ^{\epsilon }$ denote the subset of the simplex in $\Re ^{2n}$ in
which every component is at least $\epsilon $: $\Delta ^{\epsilon
}=\{((\alpha(i))_{i\in M},(\beta (j))_{j\in W}):\forall i\in M\alpha(i)\geq
\epsilon ,\forall j\in W,\beta (j)\geq \epsilon \}$. Let $S(M,W,U,V)$ denote
the set of stable matchings in the cardinal matching market $(M,W,U,V)$. In
the statement of the results below, we write the matchings in $S(M,W,U,V)$
as fractional matchings $\pi$ in which every entry in $\pi$ is either $0$ or
$1$.

\begin{prop}
\label{prop:POA} For every $\epsilon>0$, $n$, $g$, and $K > 0$ There is a
cardinal marriage market $(M,W,U,V)$, with $n$ men and women, and where
utilities $U$ and $V$ are bounded by $K$ such that
\begin{equation*}
\frac{\min_{(\alpha,\beta)\in \Delta^\epsilon} \left\{
\begin{array}{c}
\max_{\pi \in \Pi} \sum_{i=1}^n \alpha(i) \sum_{j^{\prime }=1}^n
\pi_{i,j^{\prime }} U(i,j^{\prime }) \\
+ \sum_{j=1}^n \beta(j) \sum_{i^{\prime }=1}^n \pi_{i^{\prime },j}
V(i^{\prime },j)%
\end{array}
\right\} }{ \max_{(\alpha,\beta)\in \Delta^\epsilon} \left\{
\begin{array}{c}
\max_{\pi \in S(M,W,U,V)} \sum_{i=1}^n \alpha(i) \sum_{j^{\prime }=1}^n
\pi_{i,j^{\prime }} U(i,j^{\prime }) \\
+ \sum_{j=1}^n \beta(j) \sum_{i^{\prime }= 1}^n \pi_{i^{\prime },j}
V(i^{\prime },j)%
\end{array}
\right\} } \text{ is } \Omega(n^g K)
\end{equation*}
\end{prop}

Proposition~\ref{prop:POA} shows that the gap in the sum of utilities,
between the maximizing (probabilistic) matchings, and the stable matchings,
is large and grows with the size of the market at a rate that is arbitrarily
large. Moreover, the gap is large regardless of how one weighs agents'
utilities.

Of course, the interpretation of Proposition~\ref{prop:POA} is not
completely straightforward. It does not seem right to compare the sum of
utilities in a model in which transfers are not allowed with the sum of
utilities in the TU model.\footnote{%
This problem of interpretation is present throughout the literature on the
price of anarchy.} Nevertheless, we hope that Proposition~\ref{prop:POA}
sheds additional light on the role of transfers in matching markets.

\newpage

\printbibliography

\newpage

\appendix

\section{Appendix: Proofs\label{sec:proofs}}

\subsection{Proof of Theorem \protect\ref{prop:EfficiencyMarriage}}

\begin{prf}
(i) (a) Characterization of No-trade stable matchings as in (\ref{NTS})
follows directly from the definition.

(b) Characterization (\ref{NTUStability}) follows directly from the
definition.

(c) For characterization of TU stability in terms of (\ref{TUStability}),
recall that according to the definition, matching $\sigma _{0}\left(
i\right) =i$ is TU Stable if there are vectors $u\left( i\right) $ and $%
v\left( j\right) $ such that for each $i$ and $j$,%
\begin{equation*}
u\left( i\right) +v\left( j\right) \geq U\left( i,j\right) +V\left(
i,j\right)
\end{equation*}%
with equality for $i=j$. Hence, there exists a monetary transfer $T_{i}$ (of
either sign) from man $i$ to woman $i$ at equilibrium given by
\begin{equation*}
T_{i}=u\left( i\right) -U\left( i,i\right) =V\left( i,i\right) -v\left(
i\right) .
\end{equation*}%
The stability condition rewrites as $T_{j}-T_{i}\leq R_{ij}+S_{ij}$, thus,
one is led to characterization (\ref{TUStability}).

(d) For characterization of Ex-ante Pareto efficient matchings in terms of (%
\ref{EAP}), the proof is an extension of the proof by \cite{fostel2004two},
which give in full. Assume $\sigma _{0}$ is Ex-ante efficient. Then the
Linear Programming problem
\begin{eqnarray*}
&&\max \sum_{i}x_{i}+\sum_{j}y_{j} \\
&&\text{s.t.} \\
&&x_{i}=-\sum_{j}\pi _{ij}R_{ij}\text{ and }y_{j}=-\sum_{i}\pi _{ij}S_{ij} \\
&&\sum_{k}\pi _{ik}=\sum_{k}\pi _{ki}\text{ and }\sum_{k}\pi _{ik}=1 \\
&&x_{i}\geq 0\text{, }y_{j}\geq 0\text{, }\pi _{ij}\geq 0.
\end{eqnarray*}%
is feasible and its value is zero. Thus it coincides with the value of its
dual, which is%
\begin{eqnarray*}
&&\min -\sum_{i}\phi _{i} \\
&&\text{s.t.} \\
&&v_{j}-v_{i}\leq \lambda _{i}R_{ij}+\mu _{j}S_{ij}+\phi _{i} \\
&&\lambda _{i}\geq 1\text{ and }\mu _{j}\geq 1
\end{eqnarray*}%
where variables $\lambda _{i}$, $\mu _{j}$, $v_{i}$ and $\phi _{i}$ in the
dual problem are the Lagrange multipliers associated to the four constraints
in the primal problem, and variables $\pi _{ij}$, $x_{i}$, and $y_{j}$ in
the primal problem are the Lagrange multipliers associated to the three
constraints in the dual problem. Hence the dual program is feasible, and
there exist vectors $\lambda $, $\mu $, and $\phi $, such that%
\begin{eqnarray}
&&v_{j}-v_{i}\leq \lambda _{i}R_{ij}+\mu _{j}S_{ij}+\phi _{i}
\label{firstineq} \\
&&\lambda _{i}\geq 1\text{ and }\mu _{j}\geq 1  \notag \\
&&\sum_{i}\phi _{i}=0  \notag
\end{eqnarray}%
but setting $j=i$ in inequality (\ref{firstineq}) implies (because $%
R_{ii}=S_{ii}=0$) that $\phi _{i}\geq 0$, hence as $\sum_{i}\phi _{i}=0$,
thus $\phi _{i}=0$. Therefore it exist vectors $\lambda _{i}>0$ and $\mu
_{j}>0$, such that%
\begin{equation*}
v_{j}-v_{i}\leq \lambda _{i}R_{ij}+\mu _{j}S_{ij}.
\end{equation*}

(e) For characterization of Ex-post Pareto efficient matchings in terms of (%
\ref{EPP}), assume $\sigma _{0}$ is Ex-post Pareto efficient, and let%
\begin{equation*}
Q_{ij}=\max \left( R_{ij},S_{ij}\right) ,
\end{equation*}%
so that by definition, matrix $Q_{ij}$ satisfies \textbf{\textquotedblleft
cyclical consistency\textquotedblright }: for any cycle $%
i_{1},...,i_{p+1}=i_{1}$,%
\begin{equation}
\forall k,~Q_{i_{k}i_{k+1}}\leq 0\text{ implies }\forall
k,~Q_{i_{k}i_{k+1}}=0,  \label{cyclicalConsistency}
\end{equation}%
By the Linear Programming proof of Afriat's theorem in \cite{fostel2004two}%
\footnote{%
The link between Afriat's theorem and the characterization of efficiency in
the housing problem was first made in \cite{EkelandGalichon}.}, see
implication (i) implies (ii) in \cite{EkelandGalichon}, there are scalars $%
\lambda _{i}>0$ and $v_{i}$ such that (\ref{EPP}) holds.
\end{prf}

\subsection{Proof of Theorem \protect\ref{thm:Implications}}

\begin{prf}
(i) No trade stable implies TU Stable is obtained by taking with $T_{i}=0$
in (\ref{TUStability}).

TU\ Stable implies Ex-ante Pareto is obtained by taking $v_{i}=T_{i}$ and $%
\lambda _{i}=\mu _{j}=1$ in (\ref{EAP}).

To show that Ex-ante Pareto implies Ex-post Pareto, assume there exist $%
v_{i} $ and $\lambda _{i},\mu _{j}>0$ such that $v_{j}-v_{i}\leq \lambda
_{i}R_{ij}+\mu _{j}S_{ij}$. Now assume $\max \left( R_{ij},S_{ij}\right)
\leq 0$. Then $v_{j}-v_{i}\leq 0$, and the same implication holds with
strict inequalities. By implication (iii) implies (ii) in %
\cite{EkelandGalichon}, there exist scalars $v_{i}^{\prime }$ and $\lambda
_{i}^{\prime }$ such that $v_{j}^{\prime }-v_{i}^{\prime }\leq \lambda
_{i}^{\prime }\max \left( R_{ij},S_{ij}\right) $.

No-trade stable implies NTU stable follows from $R_{ij}+S_{ij}\leq 2\max
\left( R_{ij},S_{ij}\right) $.

NTU Stable implies Ex-post Pareto is obtained by taking $\lambda _{i}=1$ and
$v_{i}=0$ in (\ref{EPP}).

Part (i) of the result is proved using a series of counterexample, which for
the most part only require two agents (one can incorporate a third neutral
agents, which has zero utility regardless of the outcome).

We show point (iii) before point (ii). In order to show (iii), it is enough
to show the following claims, proved in Examples \ref{ex1} to \ref{ex3}:

\begin{itemize}
\item TU\ Stable does not imply NTU Stable -- cf. example \ref{ex1}

\item NTU Stable does not imply Ex-Ante Pareto -- cf. example \ref{ex2}

\item Ex-Ante Pareto does not imply TU Stable -- cf. example \ref{ex3}

\item Ex-Ante Pareto does not imply NTU Stable -- cf. example \ref{ex3}

\item Ex-Post Pareto does not imply Ex-Ante Pareto -- cf. example \ref{ex2}

\item Ex-Post Pareto does not imply NTU stable -- cf. example \ref{ex3}.
\end{itemize}

To prove part (ii), we note that in the proof of part (i), the only instance
where we needed three agents was to disprove that NTU stable implies Ex-Ante
Pareto efficient and to disprove that Ex-Post Pareto efficient implies
Ex-Ante Pareto efficient. We will show that these implications actually hold
when there are only two agents. Indeed, when there are two agents, $\sigma
_{0}$ is Ex-Ante Pareto efficient if there are positive scalars $\lambda
_{1} $, $\lambda _{2}$, $\mu _{1}$ and $\mu _{2}$ such that%
\begin{equation*}
0\leq \lambda _{1}R_{12}+\lambda _{2}R_{21}+\mu _{1}S_{12}+\mu _{2}S_{21}
\end{equation*}%
which is equivalent to%
\begin{equation}
0\leq \max \left( R_{12},R_{21},S_{12},S_{21}\right) .  \label{EAP2agents}
\end{equation}

Therefore, if $\sigma _{0}$ is Ex-Post Pareto efficient, then $%
v_{j}-v_{i}\leq \lambda _{i}\max \left( R_{ij},S_{ij}\right) $. But either $%
v_{1}-v_{2}$ or $v_{2}-v_{1}$ is nonnegative, thus (\ref{EAP2agents}) holds,
and Ex-Post Pareto efficient implies Ex-Ante Pareto efficient.
\end{prf}

\subsection{Proof of Theorem~\protect\ref{thm:onematching}}

Assume $\mu _{0}\left( i\right) =i$ (this is w.l.o.g. as can always relabel
individuals). Take $R_{ij}=U\left( i,j\right) -U\left( i,i\right) $ and $%
S_{ij}=V\left( i,j\right) -V\left( j,j\right) $. $\mu _{0}$ is Stable iff $%
\min \left( R_{ij},S_{ij}\right) \leq 0$ for all $i$ and $j$, with strict
inequality for $j\neq i$. Consider%
\begin{eqnarray*}
\bar{U}\left( i,j\right) &=&\frac{1}{2}-e^{-tR_{ij}}\text{ for }i\neq j \\
\bar{U}\left( i,i\right) &=&0
\end{eqnarray*}%
one has:

\begin{itemize}
\item $\bar{U}\left( i,j\right) >0$ if and only if $\frac{1}{2}>e^{-tR_{ij}}$
that is $-\log 2>-tR_{ij}$ that is $tR_{ij}>\log 2$ hence $R_{ij}>0$.

\item $\bar{U}\left( i,j\right) <0$ if and only if $tR_{ij}<\log 2$ hence $%
R_{ij}<0$.
\end{itemize}

Take
\begin{equation*}
t>\max_{i\neq j}\left( \left\vert \frac{\log 2}{R_{ij}}\right\vert
,\left\vert \frac{\log 2}{S_{ij}}\right\vert \right)
\end{equation*}%
and let%
\begin{eqnarray*}
\bar{V}\left( i,j\right) &=&\frac{1}{2}-e^{-tS_{ij}}\text{ for }i\neq j \\
\bar{V}\left( i,i\right) &=&0
\end{eqnarray*}%
Then $\bar{U}\left( i,j\right) \leq 0$ and $\bar{V}\left( i,j\right) \leq 0$%
, thus%
\begin{equation*}
\bar{U}\left( i,j\right) +\bar{V}\left( i,j\right) \leq 0=\bar{U}\left(
i,i\right) +\bar{V}\left( j,j\right) .
\end{equation*}%
Thus $\mu _{0}$ is a No-Trade stable Matching associated to utilities $\bar{U%
}$ and $\bar{V}$. \qed

\subsection{Claim in Example~\protect\ref{ex:negative}}

Rephrasing, we want to know if there are $\alpha(m) >0$ and $\beta(w) >0$
such that, for all $(m,w)$,
\begin{equation*}
\alpha(m) (u(m) - U(m,w)) + \beta(w)(v(w)-V(m,w))\geq 0
\end{equation*}
Consider the matrix $A$ which has one column for each $m$ and each $w$, and
one row for each pair $(m,w)\in M\times W$. The matrix $A =
(a_{(m,w),a})_{(m,w)\in M\times W, a\in M\cup W} $ is defined as follows.
The row corresponding to $(m,w)$ has zeroes in all its entries except in the
columns corresponding to $m$ and $w$. It has $u(m) - U(m,w)$ in $m$'s column
and $v(w)-V(m,w)$ in $w$'s column.

The problem is to find $x\gg 0$ such that $A\cdot x \geq 0$. We introduce
the matrix $B$ such that the $i$'th row of $B$ is the vector $e_i =
(0,\ldots,1,\ldots,0)$ with a $1$ only in entry $i$. Then we want to find a
vector $x\in\Re^n$ such that $A\cdot x \geq 0$ and $B\cdot x\gg 0$. By
Motzkin's Theorem of the Alternative, such a vector $x$ exists iff there is
no $(y,z)$, with $z>0$ (meaning $z\geq 0$ and $z\neq 0$) such that
\begin{equation*}
y\cdot A + z\cdot B = 0.
\end{equation*}

\begin{equation*}
\begin{array}{cccc}
i & i^{\prime } & i_{0} & i_{1} \\ \hline
j & j^{\prime } & \underline{j} & \underline{w}^{\prime } \\
\underline{j_{0}} & \underline{j_{1}} & j_{0} & j_{0} \\
j^{\prime } & j & j_{1} & j_{1} \\
j_{1} & j_{0} & j^{\prime } & j \\
&  &  &
\end{array}%
\;\;\;\;\;%
\begin{array}{cccc}
j & j^{\prime } & j_{0} & j_{1} \\ \hline
i^{\prime } & i & \underline{i} & \underline{i^{^{\prime }}} \\
\underline{i_{0}} & \underline{i_{1}} & i_{0} & i_{0} \\
i & i^{\prime } & i_{1} & i_{1} \\
i_{1} & i_{0} & i^{\prime } & i \\
&  &  &
\end{array}%
\end{equation*}%
Utilities are:
\begin{equation*}
\begin{array}{cccc}
i & i^{\prime } & i_{0} & i_{1} \\ \hline
1+\delta & 1 & 1/2 & 1/2 \\
1/2 & 1/2 & 1/3 & 1/3 \\
0 & 0 & 1/4 & 1/4 \\
-1 & -1 & 1/5 & 1/5 \\
&  &  &
\end{array}%
\;\;\;\;\;%
\begin{array}{cccc}
j & j^{\prime } & j_{0} & j_{1} \\ \hline
1 & 1 & 1/2 & 1/2 \\
1/2 & 1/2 & 1/3 & 1/3 \\
0 & 0 & 1/4 & 1/4 \\
-1 & -1 & 1/5 & 1/5 \\
&  &  &
\end{array}%
\end{equation*}

The upper $4$ rows of $A$ are:
\begin{equation*}
\begin{array}{c|cccccccc}
& i & i^{\prime } & i_{0} & i_{1} & j & j^{\prime } & j_{0} & j_{1} \\ \hline
i,j & 1/2-(1+\delta) & 0 & 0 & 0 & 1/2 & 0 & 0 & 0 \\
i,j^{\prime } & 1/2 & 0 & 0 & 0 & 0 & 1/2-1 &  &  \\
i^{\prime },j & 0 & 1/2 & 0 & 0 & -1/2 & 0 & 0 & 0 \\
i^{\prime },j^{\prime } & 0 & -1/2 & 0 & 0 & 0 & 1/2 & 0 & 0 \\
&  &  &  &  &  &  &  &
\end{array}%
\end{equation*}%
So the sum of the first four rows of $A$ is $(-\delta,0,0,0,0,0,0,0)$.

Notice that we have other rows: for example the row corresponding to $%
(i,j_{1})$ is:
\begin{equation*}
\begin{array}{c|cccccccc}
& i & i^{\prime } & i_{0} & i_{1} & j & j^{\prime } & j_{0} & j_{1} \\ \hline
i,j_{1} & 1/2+1 & 0 & 0 & 0 & 0 & 0 & 0 & 1/2-1/5, \\
&  &  &  &  &  &  &  &
\end{array}%
\end{equation*}%
but these rows will get weight zero in the linear combination below.

So $y=(1,1,1,1,0,\ldots,0)$ and $z = (\delta,0,\ldots,0)$ exhibit a solution
to the alternative system as
\begin{equation*}
y\cdot A + z\cdot B = (-\delta,0, 0, 0 ,0,0,0,0) + \delta (1,0,\ldots, 0)= 0
\end{equation*}
\qed

Observe that in the construction of a solution to the dual system above, we
could perturb utilities by adding payoffs to each entry, in such a way that
we obtain the matrix $A+A^{\prime }$ instead of $A$ above. By choosing the
perturbation so that $y\cdot A^{\prime }=0$ the result goes through.

\subsection{Proof of Theorem~\protect\ref{prop:isolated}}

Let $S(P)$ be the set of stable matchings in the ordinal matching market $%
(M,W,P)$. Suppose that there are $N$ stable matchings, and enumerate them,
so $S(P)=\{\mu^1,\ldots , \mu^K\}$.

To prove the proposition we first establish some simple lemmas.

\begin{lemma}
\label{basic} For any $i\in M$ and $j\in W$,
\begin{equation*}
\abs{\{k : j \pm \mu^k(i) \}}+\abs{\{k : i \pw \mu^k(j) \}}\leq K-\abs{\{k :
j = \mu^k(i) \}}
\end{equation*}
\end{lemma}

\begin{proof}
Let $j>_{i}\mu^{k}(i)$; then for $\mu^{k}$ to be stable we need that $%
\mu^{k}(j)>_{j}i$. So $\left | \{k : j \mathbin{>_i} \mu^k(i) \} \right |
\leq \left | \{k : \mu^k(j) \mathbin{>_j} i \} \right | $.

Then,
\begin{align*}
\left | \{k : i \mathbin{>_j} \mu^k(j) \} \right | & =K- \left | \{k :
\mu^k(j) \mathbin{\geq_{j}} i \} \right | \\
& \leq K- \left | \{k : j \mathbin{>_i} \mu^k(i) \} \right | - \left | \{k :
j = \mu^k(i) \} \right | ,
\end{align*}%
where the last inequality follows from the previous paragraph and the fact
that preferences $>_{j}$ are strict.
\end{proof}

Let $\hat{U}(i,j)= \left | \{k : j\mathbin{\geq_{i}} \mu^k(i) \} \right | $
and $\hat{V}(i,j)= \left | \{k : i\P {j} \mu^k(j) \} \right | $. By the
previous lemma, $\hat{U}(i,j)+\hat{V}(i,j)\leq K$ for all $i$ and $j$.

\begin{lemma}
\label{isolated} If $\mu $ is an isolated stable matching, $\mu'$ is a
stable matching,  and $i,\hat{i}\in
M$, then $\mu (i)>_{{i}}\mu ^{\prime }(i)$ iff $\mu (\hat{i})>_{\hat{i}}\mu
^{\prime }(\hat{i})$.
\end{lemma}

\begin{proof}
Suppose (reasoning by contradiction) that $\mu (i)>_{i}\mu ^{\prime }(i)$
while $\mu ^{\prime }(\hat i)\geq _{\hat i}\mu (\hat i)$. Since $\mu $ is
isolated and preferences are strict, we have $\mu ^{\prime }(\hat i)>_{{\hat
i}}\mu (\hat i)$. Now let $\hat{\mu}=\mu \vee \mu ^{\prime }$, using the
join operator in the lattice of stable matchings (see \cite{roth90}). Then $%
\hat{\mu}(i)=\mu (i)$ and $\hat{\mu}(\hat i)=\mu ^{\prime }(\hat i)$. So $%
\hat{\mu}\in S(P)$, $\hat{\mu}(i)=\mu (i)$, and $\hat{\mu}\neq \mu $; a
contradiction of the hypothesis that $\mu$ is isolated.
\end{proof}

\begin{lemma}
\label{isolated2} If $\mu$ is an isolated stable matching then \[\hat
U(i,\mu(i)) + \hat V(\mu(j),j) = K.\]
\end{lemma}

\begin{proof}
We prove that
\begin{equation*}
\{k:\mu \neq \mu ^{k}\text{ and }\mu (i)\geq _{i}\mu ^{k}(i)\}=\{k:\mu \neq
\mu ^{k}\text{ and }\mu ^{k}(j)\geq _{j}\mu (j)\}.
\end{equation*}%
The lemma follows then because
\begin{align*}
\hat{U}(i,\mu (i))+\hat{V}(\mu (j),j)& = \left | \{n: \mu(i)%
\mathbin{\geq_{i}}\mu^k(i)\} \right | + \left | \{k: \mu^k(j)\P {j}\mu(j)\}
\right | \\
& =1+ \left | \{k: \mu\neq\mu^k \text{ and } \mu(i)\mathbin{\geq_{i}}%
\mu^k(i)\} \right | \\
& +(K- \left | \{k: \mu\neq\mu^k \text{ and } \mu^k(j)\mathbin{\geq_{j}}%
\mu(j)\} \right | -1).
\end{align*}

Let $\mu (i)\geq _{i}\mu ^{k}(i)$ and let $i=\mu (j)$. Since $\mu \neq \mu
^{k}$ is isolated and preferences are strict, $\mu (i)>_{{i}}\mu ^{k}(i)$.
Then by Lemma~\ref{isolated}, $\mu (i)>_{{i}}\mu ^{k}(i)$; so $j=\mu (i)$
implies that $\mu ^{k}(j)>_{{j}}\mu (j)$. Similarly, if $\mu ^{k}(j)>_{{j}%
}\mu (j)$ then $\mu (i)>_{{i}}\mu ^{k}(i)$. So $\mu (i)>_{{i}}\mu ^{k}(i)$.
\end{proof}

We are now in a position to prove the proposition.

Define a representation $U$ and $V$ of $P$ as follows. Fix $\delta $ such
that $0<\delta <1/2$. Let $U(i,j)=\hat{U}(i,j)$ and $V(i,j)=\hat{V}(i,j)$ if
there is $\mu \in S(P)$ such that $j=\mu (i)$. Otherwise, if $j$ is worse
than $i$'s partner in any stable matching, let $U(i,j)<0$ (and chosen to
respect representation of $P$); and if there is $\mu \in S(P)$ such that $%
j>_{{i}}\mu (i)$, let $\mu ^{0}$ be the best such matching for $i$, and
choose $U(i,j)$ such that $U(i,j)-U(i,\mu ^{0}(i))<\delta $. Choose $V$
similarly.

Let $\mu $ be an isolated matching. Fix a pair $(i,j)$ and suppose, wlog
that $u_{\mu }(i)-U(i,j)<0$ and $v_{\mu }(i)-V(i,j)\geq 0$ (if $u_{\mu
}(i)-U(i,j)\geq 0$ and $v_{\mu }(i)-V(i,j)\geq 0$ then there is nothing to
prove; and they cannot both be $<0$ or $(i,j)$ would constitute a blocking
pair).

First, if $i$ and $j$ are matched in some matching $\mu ^{\prime }\in S(P)$
then $u_{\mu }(i)-U(i,j)+v_{\mu }(i)-V(i,j)=u_{\mu }(i)-\hat{U}(i,j)+v_{\mu
}(i)-\hat{V}(i,j)$ so it follows that $u_{\mu }(i)-U(i,j)+v_{\mu
}(i)-V(i,j)\geq 0$ by Lemmas \ref{basic}, \ref{isolated2}, and the
definition of $\hat{U}(i,j)$ and $\hat{V}(i,j)$.

Second, let us assume that $i$ and $j$ are not matched in any matching in $%
S(P)$. Since $u_{\mu }(i)-U(i,j)<0$ we know that there is a matching that is
worse for $i$ than $j$. Let $\mu ^{0}$ be such that $j>_{{i}}\mu ^{\prime
}(i)$ implies that $\mu ^{0}(i)\geq _{i}\mu ^{\prime }(i)$. Thus $u_{\mu
^{0}}(i)-U(i,j)>-\delta $ by definition of $U(i,j)$. Since $j>_{{i}}\mu
^{0}(i)$, we also have $\mu ^{0}(j)>_{{j}}i$, or $\mu ^{0}$ would not be
stable. Then, letting $\mu ^{1}$ be the best matching in $S(P)$ for $j$, out
of those that are worse than $i$, we have $v_{\mu ^{0}}(j)-V(i,j)=v_{\mu
^{0}}(j)-v_{\mu ^{1}}(j)+v_{\mu ^{1}}(j)-V(i,j)>1-\delta $, as $\mu
^{0}(j)>_{{j}}\mu ^{1}(j)$ implies that $v_{\mu ^{0}}(j)-v_{\mu ^{1}}(j)\geq
1$ and the definition of $V(i,j)$ implies that $v_{\mu
^{1}}(j)-V(i,j)>-\delta $.

Finally,
\begin{align*}
u_{\mu }(i)-U(i,j)+v_{\mu }(i)-V(i,j)& =u_{\mu }(i)-u_{\mu ^{0}}(i)+u_{\mu
^{0}}(i)-U(i,j) \\
& +v_{\mu }(i)-v_{\mu ^{0}}(j)+v_{\mu ^{0}}(j)-V(i,j) \\
& =(u_{\mu }(i)-u_{\mu ^{0}}(i)+v_{\mu }(i)-v_{\mu ^{0}}(j)) \\
& +(u_{\mu ^{0}}(i)-U(i,j))+(v_{\mu ^{0}}(j)-V(i,j)) \\
& \geq 0+(-\delta )+(1-\delta )>0,
\end{align*}%
where the first inequality follows from the remarks in the previous
paragraphs, and from the fact that $K=u_{\mu }(i)+v_{\mu }(i)\geq u_{\mu
^{0}}(i)+v_{\mu ^{0}}(j)$ by Lemmas~\ref{basic} and~\ref{isolated2}. The
second inequality follows because $\delta <1/2$. This proves the proposition.

\subsection{Proof of Proposition~\protect\ref{prop:POA}}

Let $n$ be an even positive number. Let $(M,W,U,V)$ be a marriage market
with $n$ men and $n$ women, defined as follows. The agents ordinal
preferences are defined in the following tables:

\begin{equation*}
\begin{array}{ccccccc}
i_{1} & i_{2} & i_{3} & \cdots & i_{n-2} & i_{n-1} & i_{n} \\ \hline
j_{1} & j_{2} & j_{3} & \cdots & j_{n-2} & j_{n-1} & j_{n-1} \\
j_{2} & j_{3} & j_{4} & \cdots & j_{n-1} & j_{1} & j_{1} \\
j_{3} & j_{4} & j_{5} & \cdots & j_{n-2} & j_{1} & j_{2} \\
\vdots &  &  &  &  &  &  \\
j_{n/2} & j_{n/2+1} & j_{n/2+1} & \cdots & j_{n/2-2} & j_{n/2-1} & j_{n/2-1}
\\
&  &  &  &  &  & j_{n} \\
\vdots &  &  &  &  &  &  \\
j_{n-1} & j_{n} & j_{1} & \cdots &  &  &  \\
j_{n} & j_{1} & j_{2} & \cdots & j_{n-3} & j_{n-2} &  \\
&  &  &  &  &  &
\end{array}%
\end{equation*}%
The table means that $j_{1}$ is the most preferred partner for $i_{1}$,
followed by $j_{2}$, and so on. The women's' preferences are as follows.
\begin{equation*}
\begin{array}{ccccccc}
j_{1} & j_{2} & j_{3} & \cdots & j_{n-2} & j_{n-1} & j_{n} \\ \hline
i_{2} & i_{3} & i_{4} & \cdots & i_{n-1} & i_{1} & i_{1} \\
i_{3} & i_{4} & i_{5} & \cdots & i_{n-1} & i_{2} & i_{2} \\
\vdots &  &  &  &  &  &  \\
i_{n/2} & i_{n/2+1} & i_{n/2+2} & \cdots & i_{n/2-2} & i_{n/2-1} & i_{n/2-1}
\\
i_{n/2+1} & i_{n/2+2} & i_{n/2+3} & \cdots & i_{n/2} & i_{n/2} &  \\
\vdots &  &  &  &  &  &  \\
i_{n-1} & i_{n} & i_{1} & \cdots & i_{n-4} & i_{n-3} &  \\
i_{n} & i_{1} & i_{2} & \cdots & i_{n-3} & i_{n-1} &  \\
i_{1} & i_{2} & i_{3} & \cdots & i_{n-2} & i_{n} &  \\
i_{1} & i_{2} & i_{3} & \cdots & i_{n-2} & i_{n-1} & i_{n} \\
&  &  &  &  &  &
\end{array}%
\end{equation*}%
It is a routine matter to verify that there is a unique stable matching in
this market. It has $i_{1}$ matched to $j_{n/2}$, $i_{2}$ matched to $%
j_{n/2+1},$ and so on, until we obtain that $i_{n-1}$ is matched to $%
j_{n/2-1}$. We have $i_{n}$ matched to $j_{n}$. (The logic of this example
is that $i_{n}$ creates cycles in the man-proposing algorithm which pushes
the men down in their proposals until reaching the matching in the
\textquotedblleft middle\textquotedblright\ of their preferences; $j_{n}$
plays the same role in the woman proposing version of the algorithm).

Define agents' cardinal preferences as follows. Let
\begin{equation*}
U(i,j)=\left[ n - r_m(w) \right] \frac{1}{n^{g}}+\max \{0,n-1-r_{i}(j)\}(K-%
\frac{n-1}{n^{g}}),
\end{equation*}%
where $r_{i}(j)$ is the rank of woman $j$ in $i$' preferences. Similarly
define $V(i,j)$, replacing $r_{i}(j)$ with $r_{j}(i)$. Then, given the
preferences defined above, the agents utilities at the unique stable
matching satisfy:%
\begin{equation*}
u(i_{l})=v(j_{l})=\frac{1}{2n^{g-1}},l=1,\ldots ,n-1\text{ and }%
u(i_{n})=v(j_{n})=(n/2-1)\frac{1}{n^{g}}.
\end{equation*}%
So that the sum of all agents utilities at the unique stable matching is:
\begin{equation*}
2(n-1)(\frac{1}{2n^{g-1}})+2(n/2-1)\frac{1}{n^{g}},
\end{equation*}%
and agents' weighted sum of utilities is at most
\begin{equation*}
\max \{\frac{1}{2n^{g-1}},(n/2-1)\frac{1}{n^{g}}\}.
\end{equation*}

Consider the matchings $\mu ^{\ast }(i_{l})=j_{l}$, $l=1,\ldots ,n$, and $%
\hat{\mu}(j_{1})=i_{2}$, \ldots $\hat{\mu}(j_{n-2})=i_{n-1}$, $\hat{\mu}%
(j_{n-1})=i_{1}$, $\hat{\mu}(j_{n})=i_{n}$. Let $\pi $ be the random
matching that results from choosing $\mu ^{\ast }$ and $\hat{\mu}$ with
equal probability. Then, for all $i\neq i_{n}$ and $j\neq j_{n}$ we have
that
\begin{equation*}
\sum_{j^{\prime }}\pi _{i,j^{\prime }}U(i,j^{\prime })=\sum_{i^{\prime }}\pi
_{j,i^{\prime }}V(i^{\prime },j)=K/2,
\end{equation*}%
while
\begin{equation*}
\sum_{j^{\prime }}\pi _{i,j^{\prime }}U(i_{n},j^{\prime })=\sum_{i^{\prime
}}\pi _{j,i^{\prime }}V(i^{\prime },j_{n})=(n/2-1)\frac{1}{n^{3}}.
\end{equation*}

Then
\begin{equation*}
\sum_{i\in M}\alpha(i)\sum_{j^{\prime }\in W}\pi _{i,j^{\prime
}}U(i,j^{\prime })+\sum_{j\in W}\beta (j)\sum_{i^{\prime }\in M}\pi
_{j,i^{\prime }}V(i^{\prime },j)\geq \epsilon nK/2.
\end{equation*}%
So, regardless of the values of $\alpha$ and $\beta $ in $\Delta ^{\epsilon}$%
, the fraction
\begin{equation*}
\frac{\sum_{i\in M}\alpha(i)\sum_{j^{\prime }\in W}\pi _{i,j^{\prime
}}U(i,j^{\prime })+\sum_{j\in W}\beta (j)\sum_{i^{\prime }\in M}\pi
_{j,i^{\prime }}V(i^{\prime },j)}{\sum_{i\in M}\alpha(i)u(i)+\sum_{j\in
W}\beta (j)v(j)}
\end{equation*}%
is bounded below by
\begin{equation*}
\frac{\epsilon nK/2}{\max \{\frac{1}{2n^{g-1}},(n/2-1)\frac{1}{n^{g}}\}},
\end{equation*}%
which is $\Omega (Kn^{g})$. \qed

\end{document}